\documentclass[aps,prl,twocolumn,groupedaddress]{revtex4-1}

\usepackage[normalem]{ulem} 
\usepackage{amsmath,bm}
\usepackage{amsfonts}
\usepackage{amssymb,enumerate}
\usepackage{graphicx,hyperref}
\usepackage{amssymb}
\usepackage{epstopdf,booktabs}
\usepackage[usenames,dvipsnames]{color}
\usepackage[T1]{fontenc}
\usepackage[utf8]{inputenc}
\usepackage{natbib}
\newcommand{\atan}{\text{atg}}
\newcommand{\be}{\begin {equation}}
\newcommand{\ee}{\end {equation}}
\newcommand{\beqa}{\begin {eqnarray}}
\newcommand{\eeqa}{\end {eqnarray}}

\hypersetup{
    colorlinks,    citecolor=blue,    filecolor=blue,    linkcolor=blue,    urlcolor=blue
}

\begin{document}
\title{Determining the duration of an intense laser pulse directly in focus}
\allowdisplaybreaks
\author{Felix Mackenroth}
\email{mafelix@pks.mpg.de}
\affiliation{Max Planck Institute for the Physics of Complex Systems, Dresden, Germany}

\author{Amol R. Holkundkar}
\email{amol.holkundkar@pilani.bits-pilani.ac.in}
\affiliation{Department of Physics, Birla Institute of Technology and Science - Pilani, Rajasthan, 333031, India}
\affiliation{Max Planck Institute for the Physics of Complex Systems, Dresden, Germany}

\date{\today}
 
\begin{abstract}
We propose a novel measurement technique capable of determining the temporal duration of an intense laser pulse directly in its focus at full intensity. We show that the electromagnetic radiation pattern emitted by an electron bunch with a temporal energy chirp colliding perpendicularly with the laser pulse exhibits a distinct dependence on the pulse's duration. As the electrons emit radiation into an angular region determined by the ratio of their instantaneous energy to the laser's local field strength, the temporal change of the electrons' energy imprints information about the laser's pulse duration onto the angular radiation distribution. We quantify the interaction by a simplified analytical model and confirm this model's predictions by numerical simulations of the electrons' dynamics inside a realistically focused laser field. Based on these findings the pulse's duration can be determined to an accuracy of several percent.
% A novel mechanism is proposed to measure the pulse duration of high power laser beams by using chirped electron bunch. The electron bunch with specified energy spread is propagated perpendicular to laser propagation and polarization direction and the angular distribution of time integrated emitted radiation is calculated. It has been observed that the radiation has very well defined angular window depending on the pulse duration of the laser pulse. We believe this mechanism would be of interest to the experimentalist to measure the pulse duration of the laser pulse at the focus when high intensities are involved.  
\end{abstract}

\maketitle
\textit{Introduction -} The development of lasers towards ever higher intensities, motivated by their utility as tools for studies of fundamental physics \cite{DiPiazza_etal_2012,ELI_WhiteBook} as well as applications such as particle acceleration \cite{Esarey_etal_2009} or high-energy photon sources \cite{TaPhuoc_etal_2012,Sarri_etal_2014}, is driven by compressing their energy to ever shorter pulse durations. Lasers are now even increasingly approaching the fundamental limit of pulse durations close to only one single field oscillation with several facilities already operating in the few-cycle regime, such as the BELLA laser, compressing $39$ J of optical laser energy to pulse durations below $30$ fs \cite{Leemans_etal_2010}, the Astra Gemini laser, compressing two laser beams of $15$ J energy to pulse durations around $30$ fs \cite{Hooker_etal_2006} or the HERCULES laser, compressing an energy approaching $20$ J to pulse durations of $30$ fs \cite{Yanovsky_etal_2008}. And even more advanced facilities are already in planning, such as the SCAPA initiative, aiming at $8.75$ J of optical energy at $25$ fs pulse duration and a $5$ Hz repetition rate \cite{SCAPA}, the Petawatt Field Synthesizer, aiming at delivering $3$ J of energy within only $5$ fs at $10$ Hz repetition rate \cite{Major_etal_2009} or the Apollon 10 PW project, aiming at an energy level around $150$ J compressed to pulses as short as $15$ fs \cite{Zou_etal_2015} and several more \cite{DiPiazza_etal_2012,Kawanaka_etal_2016}. In the parameter regime disclosed by these facilities, many novel physics features are predicted to occur or were readily observed, ranging from the broadening of electron emission harmonics \cite{Akagi_etal_2016} to the dependence of atomic ionization \cite{Wittmann_etal_2009} or even electron-positron pair production \cite{Titov_etal_2016,Jansen_Mueller_2016} on the laser's sub-cycle structure. All these phenomena are predicted to feature a delicate dependence on the laser pulse's field shape, whence a thorough characterization of the laser's spatial and temporal intensity profiles is required. These profiles are conventionally characterized by collective parameters such as pulse duration, focal spot size and total pulse energy, from which then, e.g., the laser's intensity can be directly inferred \cite{Trebino_etal_1997,Walmsley_Dorrer_2009}. Conventional detector materials employed today to measure a laser's collective characteristics cannot withstand the tremendous field strengths present inside the focal volume at full intensity \cite{Trines_etal_2011}. Consequently, a laser pulse's characteristics, such as, e.g., its pulse duration \cite{Eilenberger_etal_2013}, energy \cite{PulseEnergy}, its spot size \cite{SpotSize} or even its full sub-cycle field structure \cite{Paulus_etal_2001,Paulus_etal_2003,Goulielmakis_etal_2004,Kress_etal_2006,Wittmann_etal_2009,Hoff_etal_2017} are conventionally measured either far from focus or at strongly attenuated intensity. On the other hand, laser fields may distort unpredictably upon amplification and propagation \cite{Har-Shemesh_etal_2012,Pariente_etal_2016} whence ideally a laser's characteristics should be measured under the same conditions at which an experiment is conducted. This calls for laser characterization methods operational directly in the laser's focal volume at full intensity. Thus far, however, no measurement techniques for collective pulse parameters, left alone for the laser's subcycle electric field, were implemented for intense lasers, as opposed to the recent development of corresponding techniques employed in atto-science at lower intensities \cite{Goulielmakis_etal_2004,Kueubel_etal_2017}. To overcome these shortcomings, laser pulse characterization schemes utilizing electron radiation emission patterns were suggested for operation directly in a laser's focal volume, providing measurement schemes for the laser's intensity \cite{Har-Shemesh_etal_2012} and carrier-envelope phase \cite{Mackenroth_etal_2010,Wen_etal_2012} 
with the former being already implemented experimentally \cite{Yan_etal_2017}. These characterization schemes were both based on the interaction of electrons (mass and charge $m$ and $e<0$, respectively) with initial momenta $p_i^\mu = (\varepsilon_i/c,\bm{p}_i)$, where $c$ is the speed of light, with an intense laser pulse of peak electric field $E_0$ and central frequency $\omega_0$. Here, intense refers to lasers for which the dimensionless amplitude
\begin{align}
 \xi = \frac{|e| E_0}{\omega_0 m c}
\end{align}
exceeds unity, indicating that the electrons will be accelerated to relativistic velocities within one field oscillation. A relativistic electron $\varepsilon_i \gg mc^2$, scattered from an intense laser pulse $\xi\gg1$, emits radiation into a cone around its or the laser's initial propagation direction with opening angle \cite{Harvey_etal_2009,Mackenroth_etal_2010}
\begin{align}
 \delta \zeta \sim \begin{cases}
           \frac{mc^2\xi}{\varepsilon_i} & \text{ if } \varepsilon_i \gg mc^2 \xi\\
           \frac{\varepsilon_i}{mc^2\xi} & \text{ if } \varepsilon_i \ll mc^2 \xi\\
           1 & \text{ if } \varepsilon_i \sim mc^2 \xi.
          \end{cases}
\end{align}
On the other hand, the pulse duration, an important parameter for a laser's overall characterization, cannot be determined by either of the mentioned radiation detection schemes, indicating the lack of a pulse duration measurement technique applicable in the laser's focus at full intensity.

\begin{figure}[t]
  \begin{center}
    \includegraphics[width=\linewidth]{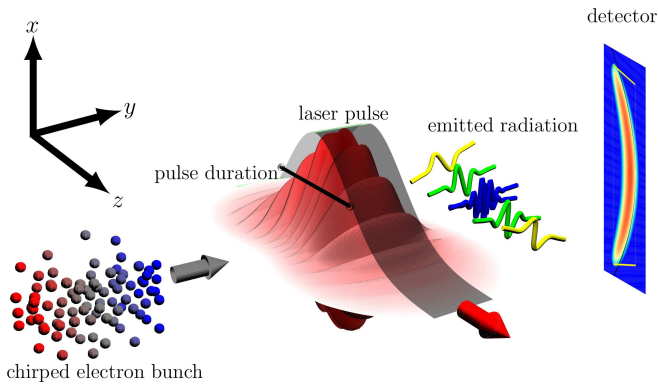}
    \caption{Schematic view of the proposed pulse duration measurement scheme and used coordinate frame.}
    \label{setup}
  \end{center}
\end{figure}

This Letter aims at filling this gap by proposing a fundamentally new approach to quantifying a laser pulse's duration directly in focus at high field strengths. Its basic working principle is to imprint a temporal structure onto an electron bunch to be scattered from the intense laser pulse, in the form of an energy chirp as are common in laser-accelerated electron bunches and need to be mitigated with great effort \cite{Brinkmann_etal_2017}. Keeping the electron bunch chirped, the particles' changing energy will lead to a temporal evolution of the angular range into which the electrons emit radiation. Here we quantify this temporal change by means of a simplified analytical model which is then benchmarked by numerical simulations, demonstrating that the pulse's duration can be quantitatively inferred. In order to make the electrons' radiation signal easily extractable from the laser's focal region and to decouple it from the strong optical laser radiation background, it is favorable to let the electrons collide with the laser pulse perpendicularly. A conceptual visualization of the proposed pulse duration measurement scheme is presented in Fig. \ref{setup}.

\textit{Theory -} In the following, we analyze the emission from an electron bunch propagating along the $y$-axis colliding perpendicularly with a laser pulse propagating along the positive $z$-axis and polarized along the $x$-axis (s.~Fig.~\ref{setup}). In order to quantify the overall angular distribution of the bunch's emitted radiation, we make use of the fact that at each time instant the radiation emitted by an ultra-relativistic electron is confined to a narrow cone around its instantaneous direction of propagation. Thus, we approximate the time-dependent direction into which radiation is emitted as the time-dependent electron propagation direction. In order to quantify the angular emission region, we thus need to solve the electrons' equation of motion, which is generally involved in a laser field of arbitrary focusing. On the other hand, it was demonstrated previously that for schemes utilizing the emission of a laser-driven electron bunch as signal it is favorable to quantify the boundaries of the angular radiation distribution and measure these in an experiment, as they provide a clear detection signal \cite{Mackenroth_etal_2010,Har-Shemesh_etal_2012}. These extremal emission angles will be determined by the strongest electron deflection, i.e., at the regions of highest field strength. Consequently, we focus on the electron dynamics close to the laser's focus, which we assume to lie in the plane $y\equiv0$ in the following. In this focal region the laser field is well approximated by a plane wave, provided the transit time of the electrons through the focal volume $\tau_t$ is longer than the pulse duration $\tau_L$, so that they will not experience non-plane wave field contributions far away from the laser focus. The ratio of these quantities, which has to exceed unity, can be found as $\tau_t / \tau_L =2w_0 / (c \tau_L)$, where $w_0$ is the laser's spot radius and we assumed the electrons to propagate with the speed of light. We thus see that the plane wave approximation is valid only for not too tight focusing or short laser pulses. It will be demonstrated, however, that for FWHM pulse durations down to $\tau_L\sim10$ fs the analytical model provides good agreement with a full numerical simulation. We assume the above condition to be satisfied and consequently approximate the electrons' dynamics by analytically known electron trajectories inside a plane wave with four-potential $A^\mu(\eta) = (m c^2 \xi / |e|) \epsilon_0^\mu g(\eta)$, with $\eta = \omega(t - z/c)$, the pulse's polarization vector $\epsilon_0^\mu$ and temporal shape $g(\eta)$. We begin by analyzing the emission angles from an electron bunch with constant energy $\varepsilon_i$. We then quantify the direction into which the electrons emit radiation in terms of the spherical coordinate angles $\theta$ and $\phi$, with respect to the $z$-axis as polar axis and the $x$-axis as azimuthal axis. As we are interested in the boundaries of the angular region into which the electrons emit radiation, we only consider the deflection of the electrons' from their initial propagation direction $\delta \zeta = \zeta - \zeta_i$ with $\zeta \in [\theta,\phi]$ and where the electrons' initial propagation direction in the considered setup is given by $\theta_i = \phi_i = \pi/2$. The deflection angles are given by the inverse tangents of the following ratios of the electrons' instantaneous momentum components $\delta \phi = \atan(p_x/p_y)$, $\delta \theta = \atan(p_z/p_\perp)$, where $p_\perp = \sqrt{p_x^2+p_y^2}$ is the electron's momentum perpendicular to the polar axis. From the analytically known expressions for an electron trajectory inside a plane wave \cite{Sarachik_Schappert_1970,Salamin_Faisal_1996} it can be derived that for the suggested perpendicular collision geometry these maximal deflection angles are given by
\begin{subequations}\label{Eq:UnchirpedAngles}
\begin{align}
 \delta \phi &= \pm \atan\left(\frac{mc \xi }{p_i}\right)\label{Eq:PhiUnchirped}\\ 
 \delta \theta &= - \atan \left(\frac{m^2 c^3 \xi^2} {2\varepsilon_i\sqrt{p_i^2 + \left(m c \xi\right)^2 }}\right)\label{Eq:ThetaUnchirped},
\end{align}
\end{subequations}
where the ambiguity in the sign of the azimuthal deflection $\delta \phi$ indicates that for the chosen setup the electrons emit symmetrically into the halfspaces $x>0$ and $x<0$, respectively. The polar deflection $\delta \theta$, on the other hand, in a plane wave is always pointed towards the laser's propagation direction, resulting in $\theta \in [\theta_i - \delta \theta , \theta_i]$. We thus see that in the $\phi$-$\theta$ plane the emission is confined to a rectangular region $(\theta,\phi) \in ([\phi_i-\delta \phi,\phi_i+\delta \phi],[\theta_i-\delta\theta,\theta_i]))$. The advantage of analyzing this \textit{emission box} rather than the angular range of only one of the angles is clearly that while the angle range in $\phi$ always depends on the ratio $mc^2 \xi/\varepsilon_i$, in the regime $\varepsilon_i \gg mc^2 \xi$ the angle range in $\theta$ depends on the square of this ratio. Consequently, the emission box is sensitive to the electron dynamics in a broader parameter range than just one of the angles.

\begin{figure}[t]
  \begin{center}
    \includegraphics[width=\linewidth]{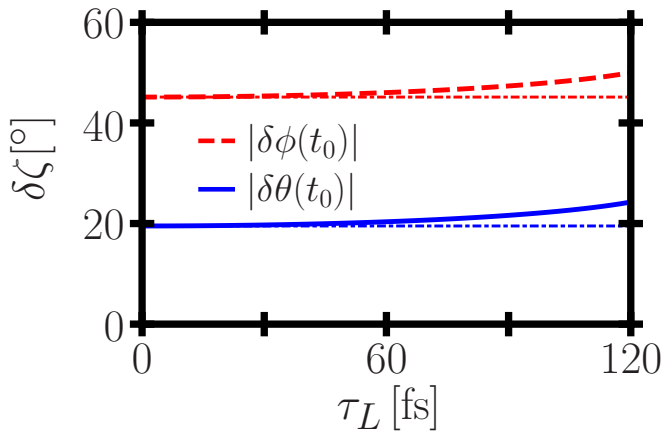}
    \caption{Theoretical predictions for the electrons' maximal angular deflections from their initial propagation direction according to Eqs.~(\ref{Eq:ChirpedAngles},\ref{Eq:ExtremalTime}) as a function of the pulse duration. For comparison we show the maximum deflection angles for an unchirped electron bunch (thin dashdotted lines).}
    \label{fig:ModelTheory}
  \end{center}
\end{figure}
We now turn to including a temporal structure in the electron bunch's energy distribution in the form of an energy chirp. Consequently, the ratio $mc^2\xi/\varepsilon_i$ will change over time, imprinting information about the pulse's temporal intensity evolution onto the temporally changing angular deflection. Naturally, as the electron bunch propagates perpendicularly to the laser pulse, there will be a different emission box for each point in space. However, we are only interested in the boundaries of the total emission box, originating from the point of strongest electron deflection. Inside a laser focus this occurs in the center of the laser focus. In our computations we chose the focus' center to coincide with the coordinate origin $(x,y,z)\equiv\bm{0}$, where we thus localize our analysis. We can then model the bunch electrons' energy to be a function only of time, which we approximate as changing linearly according to $\epsilon(t) = \varepsilon_i \left(1 - 2t/\tilde{\tau}_E\right)$, where we defined the scaled electron pulse duration $\tilde{\tau}_E := \tau_E \varepsilon_i/\Delta \varepsilon$ with the bunch's FWHM duration $\tau_E$ and its energy spread $\Delta \varepsilon$ and all other definitions as before. The negative sign for the chirp term ensures the electrons' energy to decrease over time, leading to an increase of the emission angles, avoiding the emission box's boundaries to be covered by radiation emitted at earlier times. In this case of a chirped electron bunch, the maximal emission angles will be different from the values for an unchirped electron bunch. To quantify the dynamically changing boundaries of the emission box we repeat the above analysis but now include the time dependency of both the electrons' energy as well as the laser's intensity 
\begin{subequations}\label{Eq:ChirpedAngles}
\begin{align}
 \delta \phi(t_0) &= \pm \atan\left(\frac{m c \xi g(t_0)}{p_i(t_0)}\right)\label{Eq:PhiChirped}\\ 
 \delta \theta(t_0) &= - \atan \left(\frac{m^2 c^3 \xi^2 g^2(t_0)} {2\varepsilon_i(t_0)\sqrt{p_i^2(t_0) + \left(m c \xi g(t_0)\right)^2}}\right) \label{Eq:ThetaChirped},
\end{align}
\end{subequations}
where by $t_0$ we denote the time instant at which the maximum deflection occurs. In order to determine this time instant, we assume the laser pulse to have a Gaussian shape in time $g(\eta) = \exp[-2(\eta/\tilde{\tau}_L)^2]$ with the physical and scaled FWHM pulse durations $\tau_L$ and $\tilde{\tau}_L = \tau_L/\sqrt{\log(2)}$, respectively. To now find $t_0$ we neglect the field's oscillatory structure, and compute the time at which Eqs.~(\ref{Eq:ChirpedAngles}) are extremal. It turns out, that both angles reach their maxima at
\begin{figure}[t]
  \begin{center}
    \includegraphics[width=\linewidth]{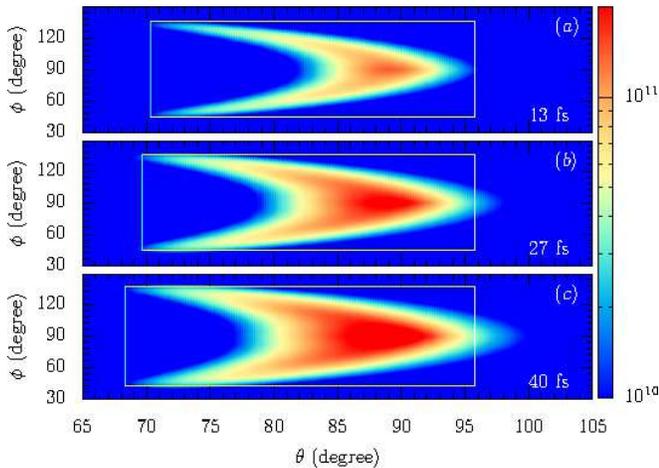}
    \caption{Numerically computed angular radiation distributions from a chirped electron bunch for different pulse durations (bottom right corners) with the analytical emission box from Eqs.~(\ref{Eq:ChirpedAngles},\ref{Eq:ExtremalTime}) superimposed (solid white rectangles).}
    \label{fig:AngleDependence}
  \end{center}
\end{figure}
\begin{align}\label{Eq:ExtremalTime}
 t_0 &= \frac{1}{2} \left(\tilde{\tau}_E  - \sqrt{\tilde{\tau}_E^2  - \tilde{\tau}_L^2 }\right).
\end{align}
We see that, due to the time dependence of the electrons' energy, contrary to the naive guess, the maxima of the emission angles, will not be reached at the field's maximum at $t = 0$, but instead $t_0$ increases for increasing pulse durations $\tau_L$. Furthermore, we see that a maximum only exists provided $\tilde{\tau}_L \leq \tilde{\tau}_E$. This condition indicates that for too long laser pulses the emission angles do not reach a maximum value but change monotonically over the whole interaction time. Consequently, as Eqs.~(\ref{Eq:ChirpedAngles},\ref{Eq:ExtremalTime}) explicitly depend on $\tau_L$, we can use the emission box's boundaries to determine the laser's pulse duration.

\textit{Results and Discussions -} To quantitatively exemplify the dependency of the emission box's boundary angles on $\tau_L$ we consider the interaction of a moderately relativistic electron bunch $\varepsilon_i/mc^2 = 10$. We assume the bunch to have a large energy spread $\Delta \varepsilon = \varepsilon_i$, as can be typical for an electron bunch from a laser-accelerator not optimized for monochromaticity \cite{Pak_etal_2010,Clayton_etal_2010}. The laser is assumed to have a dimensionless amplitude of $\xi = 10$ at a central frequency $\hbar \omega_0 = 1.55$ eV. Plotting the resulting changes of the emission box's cutoff angles (\ref{Eq:ChirpedAngles}) we find a clear dependence on the laser's pulse duration (s. Fig. \ref{fig:ModelTheory}). Thus, determining the boundary angles of the emission box emitted by a chirped electron bunch scattering from a relativistically intense laser pulse allows to determine the pulse's duration.
To corroborate the above analytical conclusions, we performed a full numerical simulation of the above scattering with a realistic laser field and electron bunch. We used the paraxial Gaussian beam model \cite{Salamin2002} for a laser pulse of not too tight focusing to a spot size of $w_0 = 10\ \mu$m. The electron bunch is assumed to be composed of 300 particles distributed according to a Gaussian density distribution in its transversal dimension and uniform in longitudinal direction, with FWHM extensions of $L = 40\ \mu$m in length and $R = 2\ \mu$m in width. To compute the radiation signal emitted by this electron bunch colliding with the given laser pulse we numerically calculate the electrons' trajectories by the relativistic Lorentz force equation of motion and calculate the resulting power radiated per unit solid angle through the standard Li$\acute{\text{e}}$nart-Wiechert potentials \cite{Jackson,Holkundkar_etal_2015,Harvey_etal_2016}. Superimposing the emission box according to Eqs.~(\ref{Eq:ChirpedAngles},\ref{Eq:ExtremalTime}) over the numerically obtained radiation signal emitted by the specified electron bunch colliding with laser pulses of various durations we find an excellent reproduction of the numerically obtained emission box boundaries by the analytical prediction (s. Fig.~\ref{fig:AngleDependence}). We thus conclude that these boundary angles of the emission box are defined by emissions close to the laser focus, where the employed plane wave model is a good approximation for the electron dynamics. We note, contrary to the analytical prediction, there occurs emission into angles $\theta > 90^\circ$, which we attribute to ponderomotive scattering from the laser focus, not accounted for in the model. To provide a qualitative estimate of this effect, we note that the ponderomotive push causes an additional angular deflection scaling as the ratio of the derivative of the electric field's square to the transverse momentum. In the present case, we can thus estimate the ponderomotive scattering angle to be of the order $\Delta \theta_\text{pond} \sim mc^2\xi^2/10^2\varepsilon_i$. Shifting the emission box boundary from $\theta = 90^\circ$ to $\theta = 90^\circ + 180^\circ \Delta \theta_\text{pond}/\pi$ in Fig.~\ref{fig:AngleDependence}, we find this estimate approximately satisfied. For all other boundary angles of the emission box the ponderomotive effect is dominated by the electron dynamics and thus not observable. We conclude that detecting the emission box cutoff angles for $\phi$ and $\theta < 90^\circ$ gives a reliable determination scheme for the laser's pulse duration.
\begin{figure}[t!]
  \begin{center}
    \includegraphics[width=\linewidth]{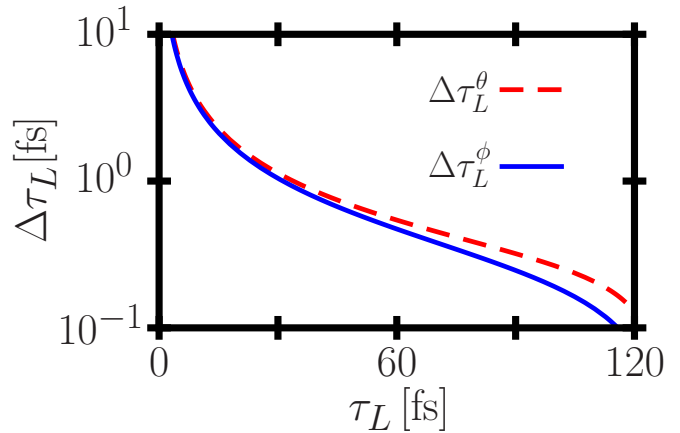}
    \caption{Error of the pulse duration measurement as a function of pulse duration resulting from a $1/10^\circ$ inaccuracy of both cutoff-angle measurements.}
    \label{fig:ErrorProp}
  \end{center}
\end{figure}
\\The proposed pulse duration scheme requires proper knowledge of the electron bunch parameters, such as its duration, energy spread, and central energy. These characteristics can most easily be obtained by spectroscopically recording the electron bunch after the interaction, when it is no longer needed. We have thus repeated the above numerical simulations including the energy loss of the electrons due to the emission of radiation in the form of radiation reaction, by replacing the Lorentz force equation by the Landau-Lifshitz equation for computing their dynamics \cite{LandauII,Mackenroth_etal_2013}. We found the difference in the radiation signal to be negligible, whence the spectral bunch properties before and after interaction with the laser pulse are comparable, facilitating a post-scattering characterization of the electron bunch.

Finally, we wish to assess the accuracy for determining the pulse duration achievable through the proposed setup. In a real experiment the cutoff angles of the emission box can be detected only up to a certain accuracy. The scheme's accuracy in determining the pulse duration, in turn, depends on the accuracy of the angle measurement. The error in the pulse duration measurement caused by an inaccuracy in the angle measurement can be derived through an error propagation to be $\Delta \tau_L^\zeta \approx \Delta \zeta (d\tau_L/d\delta\zeta)$ with $\zeta \in [\theta,\phi]$. From Eqs.~(\ref{Eq:PhiChirped},\ref{Eq:ThetaChirped}) we find complex expressions for the derivatives which we do not need to report here. Evaluating these expressions for the above studied parameters and an assumed angular accuracy of $\Delta \theta = \Delta \phi = 0.1^\circ$ we find the error of the pulse duration to be comparable for both $\theta$ and $\phi$ and to be negligible for pulse durations above $10$ fs (s. Fig.~\ref{fig:ErrorProp}). Moreover we find the error to decrease for longer pulses, as for ever longer pulses a fixed change in $\tau_L$ leads to ever increasing changes in $\delta \phi$ and $\delta \theta$.

\textit{Concluding Remarks -} We have put forward a scheme capable of determining the duration of a laser pulse of in principle arbitrarily high intensity. It takes advantage of the fact that an the electromagnetic radiation emitted by electrons interacting with an intense laser pulse is angularly confined to an \textit{emission box} depending solely on the ratio of the electron's and laser's local energy and field strength, respectively. Thus, we demonstrated that by imposing a temporal energy chirp onto the electron bunch it is possible to imprint information about the laser's temporal structure onto the integrated emission signal. This information notably also contains a clear dependence on the pulse's temporal duration, which can henceforth be measured. Assuming an angular detection accuracy of $0.1^\circ$ we have shown that the pulse duration can typically be determined with an accuracy of several percent for pulses longer than $\tau_L\gtrsim 10$ fs.

\textit{Acknowledgments -} AH acknowledges the Science and Engineering Research Board, Department of Science and Technology, Government of India for funding the project EMR/2016/002675. AH also acknowledges the local hospitality and the travel support of the Max Planck Institute for the Physics of Complex Systems, Dresden, Germany.

\end{document}